\documentclass[sn-mathphys-num]{sn-jnl}

\usepackage{graphicx}%
\usepackage{multirow}%
\usepackage{amsmath,amssymb,amsfonts}%
\usepackage{amsthm}%
\usepackage{mathrsfs}%
\usepackage[title]{appendix}%
\usepackage{xcolor}%
\usepackage{textcomp}%
\usepackage{manyfoot}%
\usepackage{booktabs}%
\usepackage{algorithm}%
\usepackage{algorithmicx}%
\usepackage{algpseudocode}%
\usepackage{listings}%
\usepackage{caption}
\usepackage{tikz}
\usetikzlibrary{shapes.geometric, arrows}
\usepackage{pgfplots}
\pgfplotsset{compat=1.15}
\usepackage{braket}
\usepackage{svg}

\graphicspath{ {images/} }

\tikzstyle{startstop} = [rectangle, rounded corners, minimum width=3cm, minimum height=1cm,text centered, draw=black, fill=red!30]
\tikzstyle{io} = [trapezium, trapezium left angle=70, trapezium right angle=110, minimum width=2.5cm, minimum height=1cm, text centered, draw=black, fill=blue!30]
\tikzstyle{process} = [rectangle, minimum width=3cm, minimum height=1cm, text centered, draw=black, fill=orange!30]
\tikzstyle{decision} = [diamond, minimum width=3cm, minimum height=1cm, text centered, draw=black, fill=green!30]
\tikzstyle{arrow} = [thick,->,>=stealth]

\theoremstyle{thmstyleone}%
\theoremstyle{thmstyletwo}%

\theoremstyle{thmstylethree}%

\begin{document}

\title{Collaborative Filtering using Variational Quantum Hopfield Associative Memory}

\author[1]{\fnm{Amir} \sur{Kermanshahani}}\email{amir.kermanshahani@piais.ir}

\author*[2]{\fnm{Ebrahim Ardeshir} \sur{Larijani}}\email{larijani@iust.ac.ir}

\author[3]{\fnm{Rakesh} \sur{Saini}}\email{rasa68842@hbku.edu.qa}

\author[3]{\fnm{Saif} \sur{Al-Kuwari}}\email{smalkuwari@hbku.edu.qa}

\affil*[1]{\orgname{Pasargad Institue of Advanced Innovative Solutions}, \city{Tehran}, \country{Iran,Islamic Republic Of }}

\affil[2]{\orgdiv{Department of Mathematics and Computer Science}, \orgname{Iran University of Science and Technology}, \orgaddress{\city{Tehran}, \country{Iran,Islamic Republic Of}}}

\affil[3]{\orgdiv{Qatar Center for Quantum Computing, College of Science and Engineering}, \orgname{Hamad Bin Khalifa University}, \orgaddress{\city{Doha}, \country{Qatar}}}

\abstract{Quantum computing, with its ability to do exponentially faster computation compared to classical systems, has found novel applications in various fields such as machine learning and recommendation systems. Quantum Machine Learning (QML), which integrates quantum computing with machine learning techniques, presents powerful new tools for data processing and pattern recognition. This paper proposes a hybrid recommendation system that combines Quantum Hopfield Associative Memory (QHAM) with deep neural networks to improve the extraction and classification on the MovieLens 1M dataset. User archetypes are clustered into multiple unique groups using the K-Means algorithm and converted into polar patterns through the encoder’s activation function. These polar patterns are then integrated into the variational QHAM-based hybrid recommendation model. The system was trained using the MSE loss over 35 epochs in an ideal environment, achieving an ROC value of 0.9795, an accuracy of 0.8841, and a F-1 Score of 0.8786. Trained with the same number of epochs in a noisy environment using a custom Qiskit AER noise model incorporating bit-flip and readout errors with the same probabilities as in real quantum hardware, achieves an ROC of 0.9177, an accuracy of 0.8013, and a F-1 Score equal to 0.7866. demonstrating consistent performance. 
Additionally, we were able to optimize the qubit overhead present in previous QHAM architectures by efficiently updating only one random targeted qubit. This research presents a novel framework that combines variational quantum computing with deep learning, capable of dealing with real-world datasets with comparable performance compared to purely classical counterparts. Additionally, the model can perform similarly well in the noisy configurations, showcasing a steady performance and proposing a promising direction for future usage in recommendation systems.}




\maketitle

\section{Introduction}\label{sec1}

The rapid progress of quantum computing has created new opportunities in artificial intelligence, especially in machine learning. Classical machine learning methods are effective but struggle with complex datasets. Quantum algorithms by utilizing principles of quantum mechanics, have shown the potential to significantly improve data processing capabilities, leading to transformative applications in areas like drug discovery and optimization \cite{1}.

Recent advancements in quantum algorithms, including the Variational Quantum Eigensolver (VQE) and the Quantum Approximate Optimization Algorithm (QAOA), illustrate the innovative methods being explored to address complex problems via hybrid quantum-classical computing frameworks. \cite{2},\cite{3}. These advancements not only address computational challenges and optimization but also serve as a foundation for exploring quantum advantages in future computational tasks.

Quantum memories have become essential in quantum information processing, enabling the effective storage and retrieval of quantum information. Early theoretical explorations laid the groundwork for practical quantum memory devices, with experimental efforts demonstrating the feasibility of quantum storage techniques across various physical systems \cite{4}, \cite{5}. Advancements in quantum memory, such as dynamical decoupling and error correction techniques, have greatly enhanced the performance and reliability of these systems. \cite{6},\cite{7}.

Additionally, the convergence of quantum mechanics and neural network paradigms has resulted in the advancement of Quantum Hopfield Networks (QHNs), which utilize quantum principles to perform tasks like associative memory and optimization more effectively. \cite{8}. Recent studies have concentrated on hybrid quantum-classical architectures that combine quantum memories and QHNs, increasing their potential applications in machine learning, optimization, and cryptography\cite{9}.

On the other hand, Transfer learning became an imperative method for machine learning by allowing models trained for one task to be effectively adapted for similar tasks, thereby reducing the need for large datasets for specific tasks. \cite{10}. Moreover, Quantum Transfer Learning (QTL) combines the advantages of quantum variational computing with classical transfer learning methods aiming to improve learning efficiency in hybrid quantum-classical learning models\cite {11}.    

Recently, novel quantum algorithms have been introduced for recommendation systems. In particular, the Quantum Collaborative Filtering (QCF) recommendation algorithm illustrates the potential of quantum computing to improve collaborative filtering techniques, demonstrating the relevance of using quantum computing in personalized recommendation tasks. \cite{12}. Quantum context-aware recommendation systems utilizing tensor singular value decomposition have emerged, leading to enhanced recommendations that take contextual information into account. \cite{13}. Moreover, research on quantum contextual bandits highlights the advantages of integrating this technique in recommendation systems for quantum data \cite{14}.

The crux variational quantum algorithm use Quantum Variational Circuits (QVCs) \cite{15} in combination with classical learning primitives. This, in turn, opens up a testbed to examine the applicability of variational hybrid quantum-classical schemes across various fields. 

This paper proposes a framework that integrates a classical autoencoder neural network with a variational quantum Hopfield associative memory, that uses quantum transfer learning. The goal of this hybrid classical-quantum network is to explore its potential for real-world applications at scale, in recommendation systems, in contrast to the current works based on matrix factorization, which suffer from scalability problems. To our understanding, this is the first work that uses quantum associative memory for recommendation systems, incorporating an industrial-scale dataset.

Section II discusses the methodology used throughout our research: how we processed the raw dataset, the model architecture, and the specifications of the QHAM variational circuit. In Section III we present the results of training our hybrid model and will make a comparison with the existing works.

\section{General Framework}\label{sec2}
This section describes the main components of our paper, 
including the learning model architecture, dataset specifications, and preprocessing procedures. By integrating a deep autoencoder neural network with K-Means clustering to extract users' archetypes and utilizing a variational circuit for implementing the Quantum Hopfield Associative Memory (QHAM), in order to memorize those archetype patterns, this section presents a novel technique for a recommendation system that utilizes quantum transfer learning within a hybrid quantum-classical network. Furthermore, some implementation details will be discussed.

\subsection{Model Architecture}
Here, we outline the architecture of the model that we developed for the hybrid quantum-classical recommendation system. We describe the structure of the deep autoencoder neural network used for dimensionality reduction, as well as the hybrid quantum-classical network that incorporates quantum Hopfield associative memory and is integrated with the autoencoder. We discuss the design choices, parameter settings, and the rationale behind selecting specific components of the model to provide a comprehensive understanding of the system's architecture.

\subsubsection{Autoencoder Neural Network}
\begin{figure}[h!]
    \centering
    \includegraphics[width=0.8\textwidth]{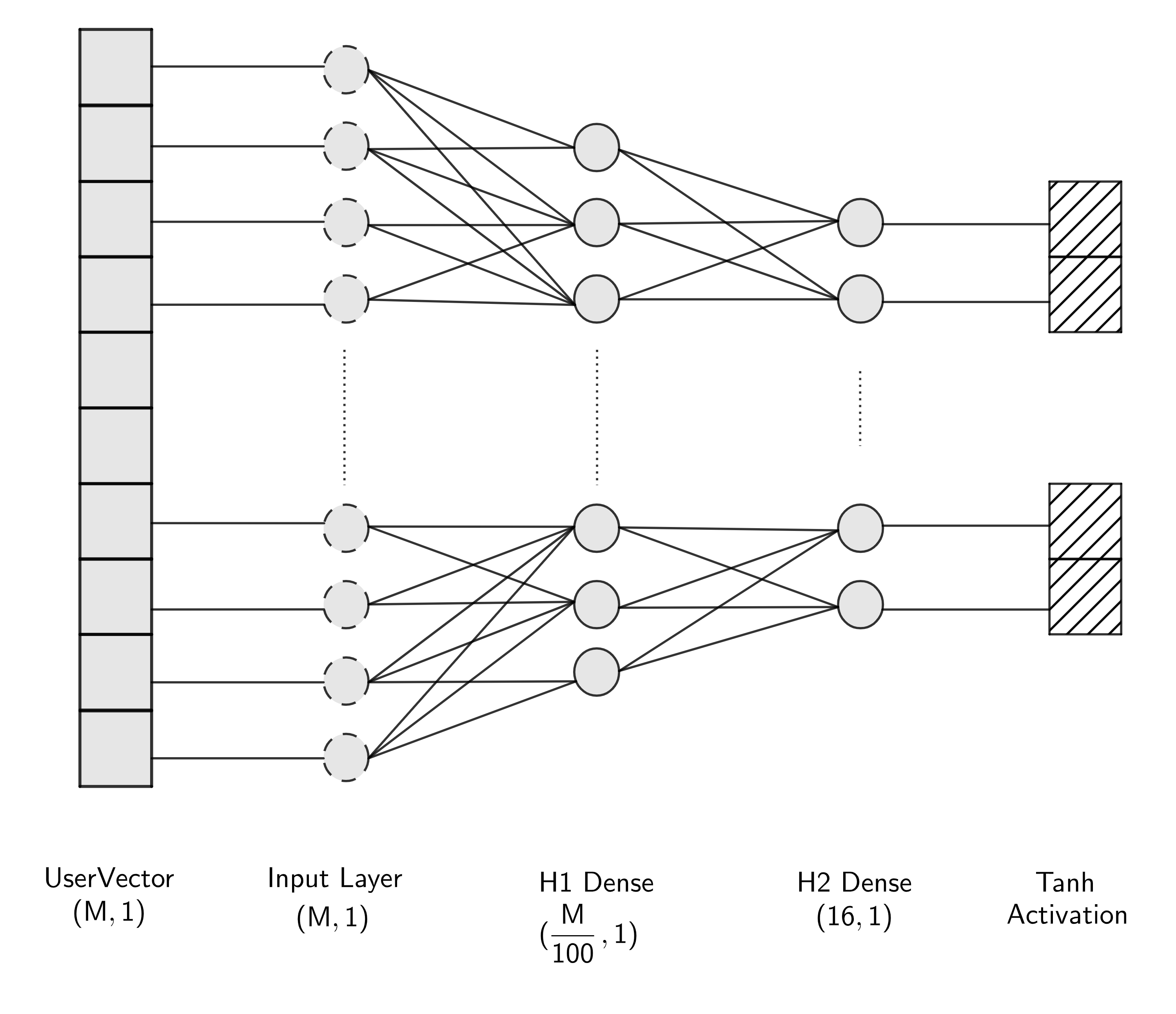}
    \captionsetup{justification=centering}
    \caption{Encoder Architecture}
    \label{fig:encoder-architecture}
\end{figure}

The encoder architecture consists of an input layer, two hidden layers, and an output layer with a hyperbolic Tangent
(Tanh) as an activation function. The input layer consists of \(M\) neurons, each corresponding to a user's rating for a specific movie, where \(M\) is the number of movies in each user ratings vector. These ratings range from 0.0 to 5.0 with increments of 0.5 and are normalized using a min-max algorithm before being passed to the network.

The first hidden layer is a dense layer with \(M/100\) neurons, designed to accommodate the input data without needing an activation function due to the relatively small value range while preserving all the information.

The second hidden layer is another dense layer that uses a "Tanh" activation function. This activation function adds non-linearity to the model, enabling it to capture complex relationships between the input and hidden layers. It projects values slightly above -1 and below +1, which are then passed to the quantum network.

The encoder compresses user rating vectors into a smaller dimension, enabling their use by the quantum Hopfield network without concerns about the number of qubits or data loss. The selection of the number of qubits is based on experimental analysis, which evaluates how effectively user patterns can be distinguished. This choice also considers the capacity of the quantum Hopfield associative memory to efficiently search through these patterns to identify the most similar one among the stored options. 

\subsubsection{Quantum Hopfield Associative Memory} 
There are different embedding methods for converting classical data points to quantum states, one of which is amplitude embedding. This is the widely adopted method in applications containing quantum machine learning besides phase embeddings. As discussed in \cite{16}, each classical neuron state \(x_i\) in a quantum system, can be mapped to a qubit that is in either the state \(\vert0\rangle\) or \(\vert1\rangle\). We associate each of these neuron states with a probability of measuring \(\vert1\rangle\) in the qubit. As a result, \(x_i = -1\) corresponds to a pure \(\vert0\rangle\) state (with \(P\vert1\rangle = 0\)), and \(x_i = 1\) corresponds to a pure \(\vert1\rangle\) state (with \(P\vert1\rangle = 1\)). 

Any non-classical value of \(x_i\) that falls within the interval \((-1, 1)\) can also be represented as a superposition state. For instance, \(x_i = 0\) is represented by the state \(\ket {s_i} = \frac{\ket0 + \ket1}{\sqrt{2}}\), which corresponds to \(P\vert1\rangle = 0.5\). 

The mapping of classical states to quantum states in this manner is defined by equation \ref{quantum_state_preparation}:
    
\begin{equation}\label{quantum_state_preparation}
    \ket{s_i} = cos(x_i\frac{\pi}{4} + \frac{\pi}{4})\ket{0} + sin(x_i\frac{\pi}{4} + \frac{\pi}{4})\ket1
\end{equation}\\
    
The full quantum system is thus represented by:

\begin{equation}\label{quantum_vector}
    \ket s=\ket{s_1s_2...s_n}
\end{equation}

\begin{figure*}[h!]
    \centering
    \includegraphics[width=0.8\textwidth]{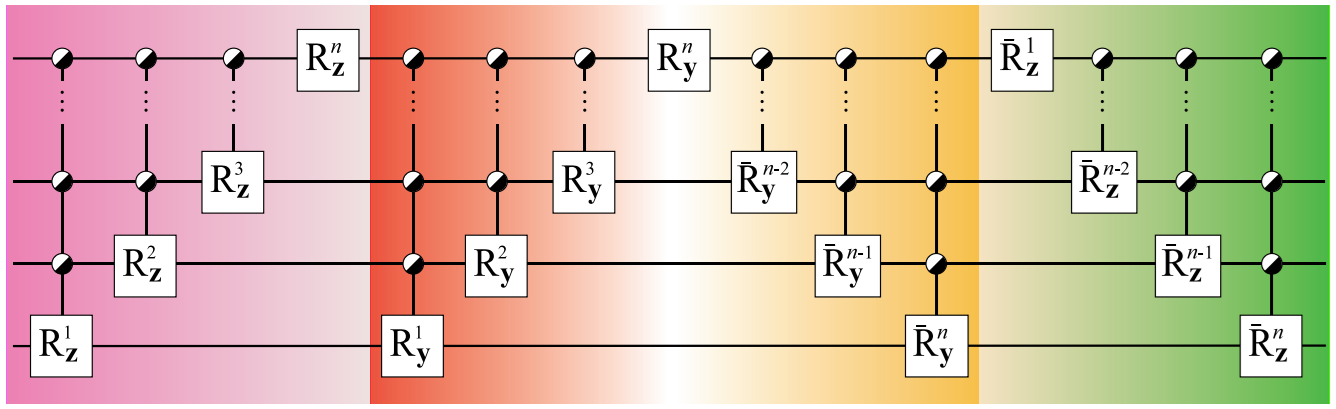}
        \caption{Mottonen State Preparation Diagram from Mottonen \emph{et al.} \cite{22}}
    \label{fig:mottonen-state-preparation}
\end{figure*}

The Mottonen State Preparation algorithm \cite{22} is employed to transform a qubit register that is initialized in the \(\ket{0}^{\bigotimes n}\) state into the desired quantum state \(\ket{s}\). This transformation is achieved by implementing equation \ref{quantum_state_preparation} for each qubit present in the quantum system, which involves systematically applying a sequence of quantum gates. The specific values of the rotation angles are derived from the target state using Mottonen’s algorithm. We utilized the existing function for implementing the Mottonen state preparation algorithm within the PennyLane framework \cite{24}. A schematic diagram of this algorithm is presented in Fig. \ref{fig:mottonen-state-preparation}.

\begin{figure}[h]
\centering
\includegraphics[width=0.8\textwidth]{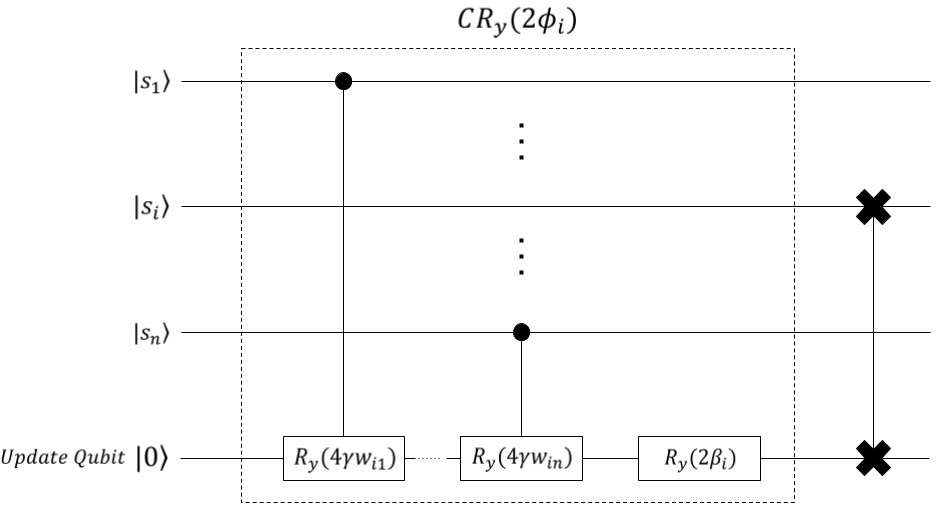}
\caption{Quantum circuit ansatz used in this paper to simulate the classical neuron. It is implemented by a series of controlled \(R_y\) rotations followed by another \(R_y\) rotation and a \(SWAP\) gate as described in \cite{16}. The angles are controlled by the weights matrix and bias vector.} 
\label{fig:quantum-neuron}
\end{figure}

The quantum neuron is implemented using an enhanced version of the framework described in the work by Miller \emph{et al.}.\cite{16}. The circuit design is illustrated in Fig~\ref{fig:quantum-neuron}. We set the number of updates to one and employed a variational form to train the circuit parameters, allowing us to achieve the same results without needing multiple updates, which would increase the qubit overhead in the system.

The model stores polar patterns of user archetypes as attractors within the quantum state space. This is done using the calculated weights matrix, as described by equation \ref{weights_matrix}. The probability of measuring \(\vert1\rangle\) at the output qubit is equal to \(sin^2(\phi)\). This output probability is the same as obtained by applying a rotation of \(CR_y(2\phi)\). The parameter \(\phi\) is related to the classical \(\theta\) parameter, as shown in equation \ref{phi-calculation}.

This output probability functions similarly to a classical neuron, where for \(\phi > \frac{\pi}{4}\), the probability \(P(\vert1\rangle)\) approaches \(1\), and for \(\phi < \frac{\pi}{4}\), \(P(\vert1\rangle)\) approaches \(0\). This behavior corresponds to the classical states \(x_i = \pm1\). Here, \(\gamma\) serves as a normalization factor, with the constraint \(0 < \phi < \frac{\pi}{4}\). This configuration allows us to access the threshold behavior of Hopfield associative memory, specifically when \(h_i = 0\). The mathematical relation for calculating \(\gamma\) is detailed in Equation~\ref{gamma_calculation}.

The \(CR_y(2\phi_i)\) rotation gate consists of rotations of the neuron state \(s_i\), which are controlled by each \(s_j\) other than the target \(s_i\) qubit. These rotations depend on the elements of the neural weight matrix \(w_{ij}\). Following these rotations, an additional rotation is implemented using the bias term value~\(\beta\), as described in equation \ref{beta_calculation}.

\begin{equation}\label{weights_matrix}
        w_{ij} = \frac{1}{m}\sum_{\mu=1}^{m}{\epsilon_{i}^{\mu} \epsilon_{j}^{\mu}}
\end{equation}

\begin{equation}\label{gamma_calculation}
        \gamma = \frac{\frac{\pi}{4}}{w_{max}^{n}}
\end{equation}

\begin{equation}\label{beta_calculation}
        \beta_i = \frac{\pi}{4} - \gamma(\sum_{j=1}^{n}{w_{ij}})
\end{equation}

\begin{equation}
\label{phi-calculation}
    \phi = \gamma\theta + \frac{\pi}{4}
\end{equation}

To update an input qubit state \(\ket{s_i}\) to a new state \(\ket{ s_i'}\), we begin by performing a controlled rotation of \(2\phi\) on the ancilla qubit, which is initialized to the state \(\ket{0}\). After this operation, the output from the update qubit is swapped with the target input qubit using a quantum SWAP gate. This results in the input qubit being updated to approach the target attractor, which either increases or decreases the probability of it being in the state \(\ket{1}\). The qubit used for the update is not recycled to avoid the reset operations and mid-circuit measurements mentioned in the work of Cao et al. \cite{23}. This approach enhances the accuracy of the neuron and allows for implementation in hardware that does not support reset operations.

Using the mentioned set of quantum gate operations, the network effectively searches for the archetypal polar pattern that best corresponds to the currently active encoded input pattern. It outputs the expectation value of the Pauli-Z measurements for the qubits that represent the active user pattern.

\begin{figure*}[h!]
    \centering
 \includegraphics[width=1\textwidth]{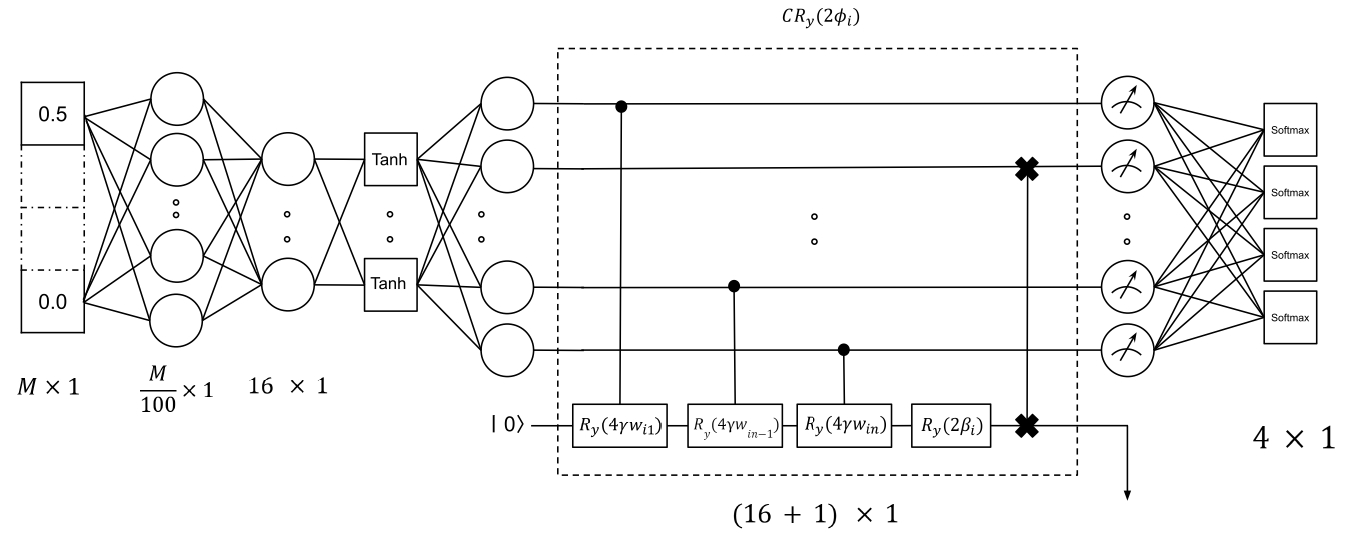}  
    \caption{Hybrid Model Architecture. It Combines the encoder deep neural network for dimension reduction and the QHAM variational circuit to obtain the stored pattern and a final linear layer to decide the user archetype category.}
    \label{fig:hybrid-model-architecture}
\end{figure*}

The output from the QHAM network is then processed by a classical neural network, which includes a dense layer followed by a SoftMax activation function. This setup produces a one-hot encoded pattern corresponding to the retrieved user archetype. The schematic representation of the entire network architecture is illustrated in Fig. \ref{fig:hybrid-model-architecture}.

\subsection{Implementation Details}
In this section, we outline the details of our implementation procedure. This includes a description of the dataset specifications, the evaluation metrics utilized, and the computational resources employed. We also discuss the quantum simulators and the noise model used to train the hybrid network.

\subsubsection{Dataset Specifications and Preprocessing}
We leveraged the MovieLens dataset, specifically the "ratings" subset of the MovieLens 1M dataset. Widely regarded as a benchmark in the field of recommendation systems, the MovieLens dataset contains user ratings and movie metadata essential for research and analysis. This dataset, including its "ratings" subset, is publicly available for usage from MovieLens.\footnote{
at https://grouplens.org/datasets/movielens/1m/}

The “ratings” subset of the chosen MovieLens dataset includes 1 million ratings for 3,706 movies, provided by 6,040 users. It is organized with user IDs, movie IDs, ratings, and timestamps. This subset captures the complex interactions between users and movies, which are essential for collaborative filtering and recommendation algorithms.

Each row in the dataset represents an individual rating given by a user for a specific movie, along with the corresponding timestamp. Each user in the "ratings" subset is uniquely identified by a user ID, which serves as a distinct marker to differentiate and track individual user preferences and behaviors within the dataset.

The main element of the “ratings” subset consists of user ratings for various movies. These ratings are assigned on a scale ranging from 0.0 to 5.0, with increments of 0.5. Each rating represents a user's evaluation of a movie's quality or enjoyment, providing essential feedback for recommendation algorithms.

The timestamps next to each rating indicate when the rating was recorded. These timestamps can help analyze user behavior over time and identify trends and patterns in movie preferences. However, for collaborative filtering, we will not use these timestamps since our focus is solely on the interactions between users and movies, irrespective of when the ratings were submitted.

The architecture of the “ratings” subset can be understood as a sparse matrix, where each row represents an individual user, each column corresponds to a movie, and each cell contains a specific user's rating for that movie along with the associated timestamp. This matrix captures the rich interactions between users and items in the dataset, forming the basis for collaborative filtering and recommendation generation techniques. We used all the users' data for the simulation on the ideal and noisy environments, with the requirement that each user had rated at least 20 movies.

We divided each preprocessed dataset into training and test-validation subsets using a 0.33 ratio. Next, we split the test-validation subset into validation and test subsets using the same ratio. 

The k-means clustering algorithm segments users into archetypes based on their preferences. This process aims to identify distinct user groups within the dataset, allowing the recommendation system to efficiently find the most similar archetype for the active user input.

The sparse nature of the rating dataset presents challenges in processing it effectively. One key challenge is extracting relevant and meaningful information from the high-dimensional, sparse space of movie ratings and transforming it into a low-dimensional dataset while preserving data integrity. We also need to clean empty ratings without losing valuable information. 

To tackle these challenges, we applied a mean-max normalization technique that scales user rating values to a range between 0 and 1. Furthermore, we transformed the original high-dimensional user ratings into a smaller vector space using our encoder neural network, based on the architecture outlined in Fig \ref{fig:encoder-architecture}. This approach ensures that no data is lost during the transformation process.

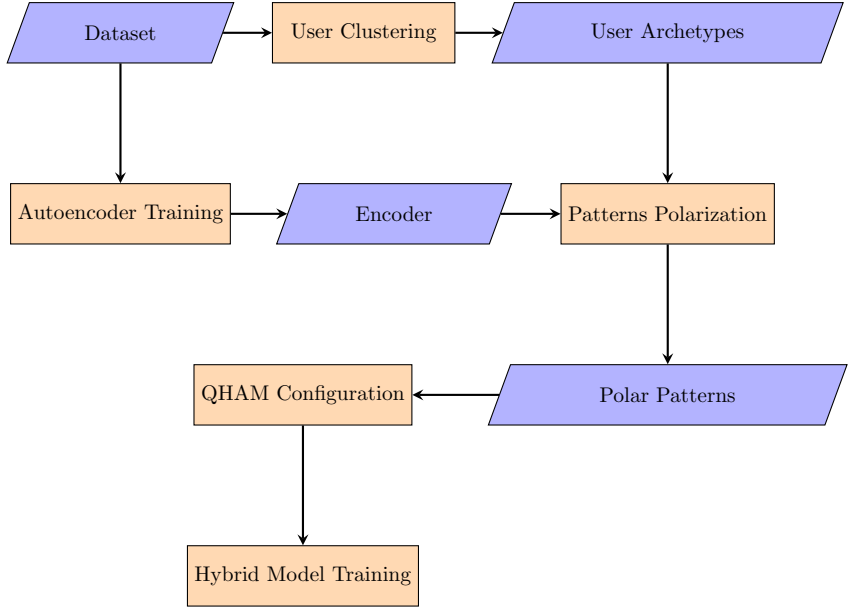
\begin{figure}
\centering
\begin{tikzpicture}[node distance=2cm, scale=0.8, transform shape]
    \centering
    \node (dataset) [io] {Dataset};
    \node (clustering) [process, right of=dataset, xshift=2cm] {User Clustering};
    \node (user_archetypes) [io, right of=clustering, xshift=3cm] {User Archetypes};
    \node (autoencoder_training) [process, below of=dataset, yshift=-1cm] {Autoencoder Training};
    \node (encoder) [io, right of=autoencoder_training, xshift=2.5cm] {Encoder};
    \node (pattern_polarization) [process, right of=encoder, xshift=2.5cm] {Patterns Polarization};
    \node (polar_patterns) [io, below of=pattern_polarization, yshift=-1cm] {Polar Patterns};
    \node (qham_config) [process, left of=polar_patterns, xshift=-4cm] {QHAM Configuration};
    \node (hybrid_model_training) [process, below of=qham_config, yshift=-1cm] {Hybrid Model Training};

    \draw [arrow] (dataset) -- (clustering);
    \draw [arrow] (clustering) -- (user_archetypes);
    \draw [arrow] (dataset) -- (autoencoder_training);
    \draw [arrow] (autoencoder_training) -- (encoder);
    \draw [arrow] (encoder) -- (pattern_polarization);
    \draw [arrow] (user_archetypes) -- (pattern_polarization);
    \draw [arrow] (pattern_polarization) -- (polar_patterns);
    \draw [arrow] (polar_patterns) -- (qham_config);
    \draw [arrow] (qham_config) -- (hybrid_model_training);

\end{tikzpicture}
\caption{Model training scheme}
\label{fig:implementation-flowchart}
\end{figure}

The overall training procedure in our model is shown schematically in Fig~\ref{fig:implementation-flowchart}.

\subsubsection{Evaluation Metrics}
To evaluate the performance of the developed recommendation system, we use the ROC, F1-score, and accuracy metrics, which are commonly employed metrics in the field and 
provide a comprehensive analysis of the model's quantitative performance.

The F1-Score is a crucial metric for evaluating the balance between precision and recall in a classification task. It is the harmonic mean of precision and recall, ensuring that both false positives and false negatives are accounted for in the model’s performance. A higher F1-Score indicates a well-balanced classifier, particularly in cases where class distributions are imbalanced. This metric provides a reliable assessment of the recommendation system’s ability to correctly identify relevant items while minimizing incorrect predictions.

The Accuracy metric measures the overall proportion of correctly classified instances out of all predictions made by the model. While simple and intuitive, accuracy may not always provide a complete picture, especially in datasets with class imbalances. However, in balanced settings, it serves as an effective benchmark for evaluating the recommendation system’s general performance. By examining accuracy alongside other metrics, we can assess how well the model aligns with real-world recommendation needs.

The Receiver Operating Characteristic (ROC) Curve illustrates the trade-off between the true positive rate (recall) and the false positive rate across different classification thresholds. The ROC score provides a comprehensive evaluation of the model’s ability to distinguish between positive and negative instances. A higher ROC score signifies better discrimination capability, with a value of 1 representing perfect classification and 0.5 indicating random performance. This metric is particularly useful in assessing the model’s ranking ability in recommendation scenarios where decision thresholds may vary.

These evaluation metrics are computed on the test set to ensure an unbiased assessment of the system’s predictive performance. Analyzing their values across different user groups provides deeper insights into the strengths and limitations of our approach. By comparing these results, we can identify areas for improvement and refine the model to enhance its recommendation accuracy and reliability.

\subsubsection{Computational Resources}
The development and evaluation of the recommendation system were carried out using classical hardware. It was implemented in Python, utilizing the PennyLane library to provide the necessary tools for building and training a hybrid deep neural network. Additionally, the quantum component was simulated using the PennyLane framework, which allows for the implementation of quantum circuits and the execution of quantum algorithms in both ideal and noisy environments. 

All experiments were conducted on a standard workstation with the following specifications:

- GPU: 3× NVIDIA RTX 6000 Ada – 48GB GDDR6 memory, 18,176 CUDA cores, 568 Tensor cores, 300W

- CPU: AMD Ryzen Threadripper PRO 7995WX – 96 cores, 2.5~5.1GHz, 480MB cache, PCIe 5.0

- System Memory: 512GB DDR5-4800

\subsubsection{Noise Model}
In order to introduce noise into our simulation to an extend which can mimick the real-world hardware situation efficiently, we followed the model present in \cite{28}. It identifies three main sources of errors in quantum circuits:
\begin{enumerate}
    \item Qubit Instability: Qubits lose their excited states over time due to relaxation and dephasing mechanisms, leading to errors in state stability and
    an increase in circuit execution time.
    \item Gate Errors: Quantum gates introduce inaccuracies during operations. Single-qubit gates typically have error probabilities between \(10^{-4}\) and \(10^{-3}\), while two-qubit gates have higher error probabilities ranging from  \(10^{-3}\) to \(10^{-2}\). 
    \item Measurement Errors: Errors occur during the physical measurement of qubit states, with probabilities around \(10^{-2}\).
\end{enumerate}

We configured the noisy environment using the Qiskit AER noise model \cite{27} and introduced single-qubit and two-qubit Bit-Flip errors, affecting the state with a \(\sigma_x\) matrix, on 6 randomly selected gates with sampled probabilities within the range of 0.001 to 0.01. We also introduced readout errors on 6 randomly selected measurement operations with a probability sample within the range of 0.01 to 0.07. The range of the probabilities is chosen so that they are slightly higher than the noise probabilities present in real quantum hardware

\section{Results}\label{sec3}
The results section presents the evaluation of our model, benchmarked against established methods and datasets. The evaluation details of our model are organized into three key stages: autoencoder training, user archetype extraction, and hybrid model training.

\subsection{Autoencoder Training}

\begin{figure}
    \centering
    \includegraphics[width=0.8\textwidth]{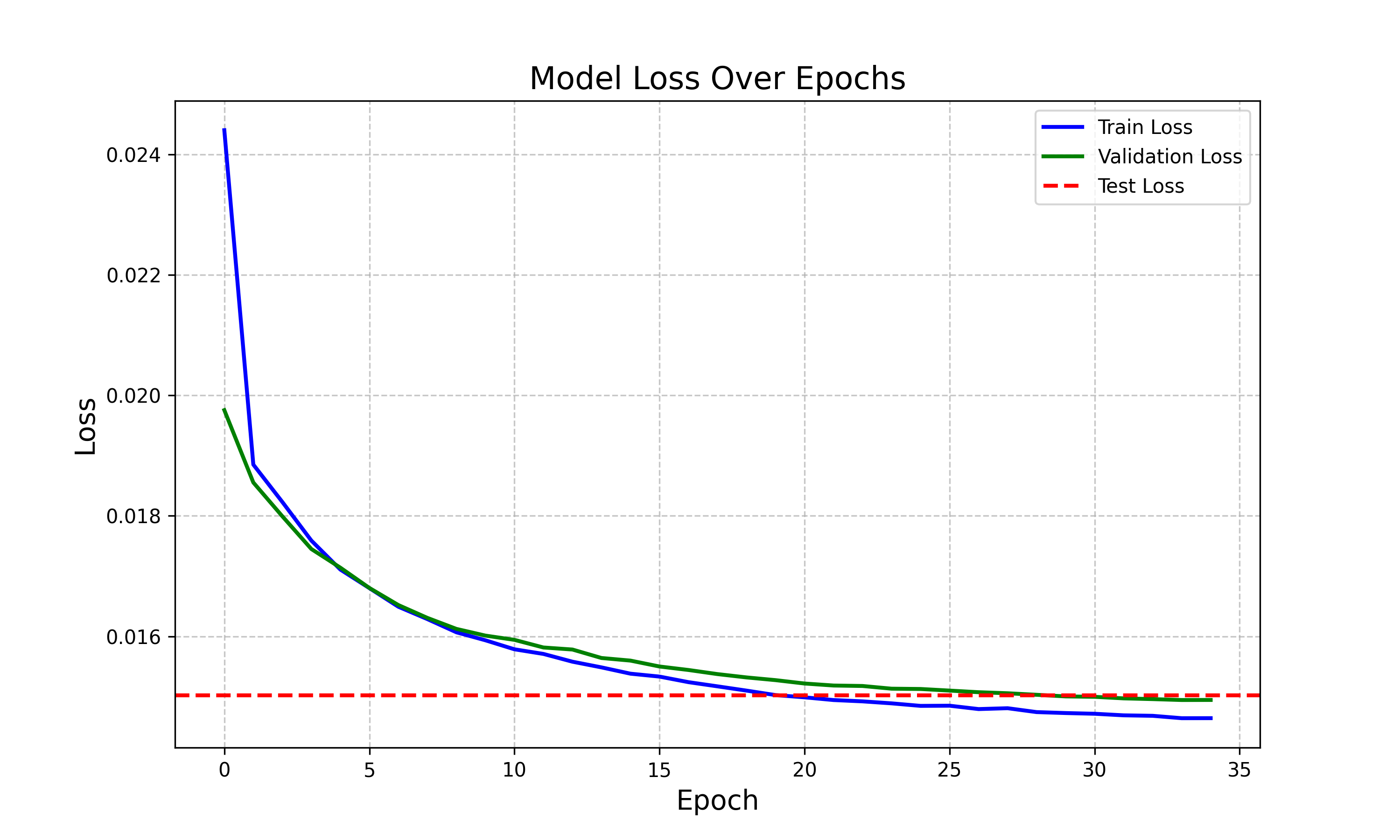}
    \caption{Autoencoder training results}\label{auto_training}
\end{figure}

We trained the autoencoder using the user-item interaction matrix derived from the MovieLens datasets, as described in Section \ref{sec2}. The autoencoder compressed the high-dimensional user-item data into a small latent space. The training process lasted for 35 epochs, and the mean squared error (MSE) on the test dataset reached 0.015. You can see the plot for the training phase as well as the final MSE value on the test set in Fig~\ref{auto_training}.

\subsection{User Archetype Extraction}
\begin{figure}[h!]
    \centering    \includegraphics[width=0.6\textwidth]{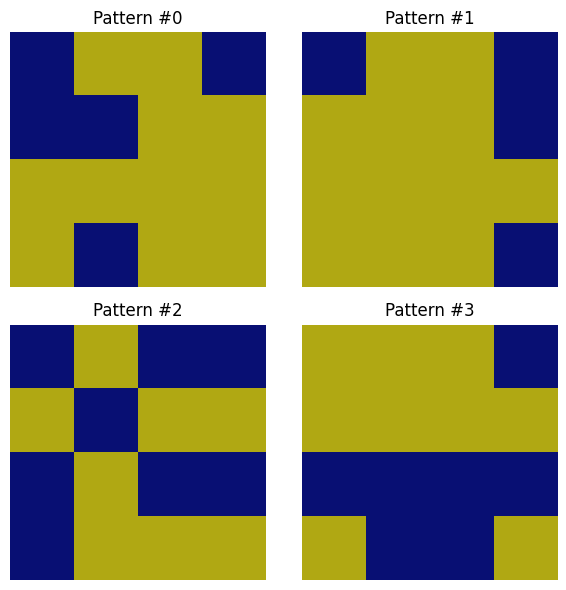}
    \caption{User Archetype Patterns}
    \label{fig:archetype-patterns}
\end{figure}

Using the K-Means clustering algorithm, we identified user archetypes from the preprocessed dataset. This resulted in four distinct user groups, representing diverse movie preference patterns. The cluster centroids were transformed into binary patterns through the encoder component of the autoencoder, which utilized a polar activation function. These archetype patterns are illustrated in Fig \ref{fig:archetype-patterns}.

\subsection{Hybrid Model Performance}
The polar user archetypes were then used to calculate the weights matrix and bias vector for the Quantum Hopfield Associative Memory (QHAM) network, following Equations~\ref{weights_matrix} and \ref{beta_calculation}.

We implemented our model using the PennyLane library, running it on both ideal simulators that perform matrix algebra for the circuit and in a noisy environment. Specifically, we utilized the IBM Qiskit \cite{25} aer simulator back-end \cite{26} and introduced noise to the system using the aer noise models.

Table \ref{table:hybrid-model-results} summarizes the hybrid model's performance metrics on the test set for both ideal and noisy conditions.

\renewcommand{\arraystretch}{1.5}  
\begin{figure*}
    \centering
    \includegraphics[width=\textwidth]{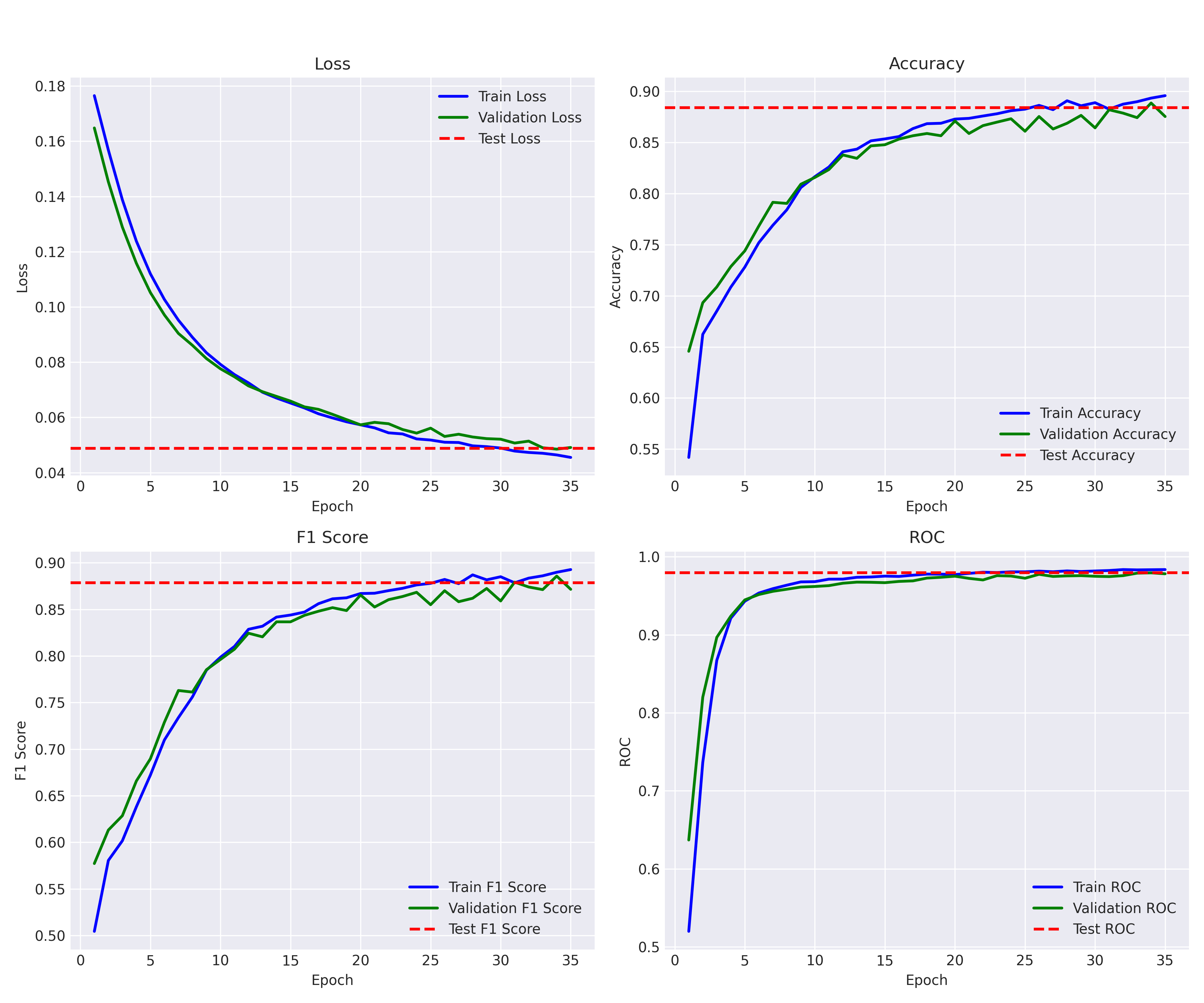}
    \caption{Model performance in ideal environment}\label{fig:ideal-simulation-plot}
\end{figure*}

\begin{figure*}
    \centering
    \includegraphics[width=\textwidth]{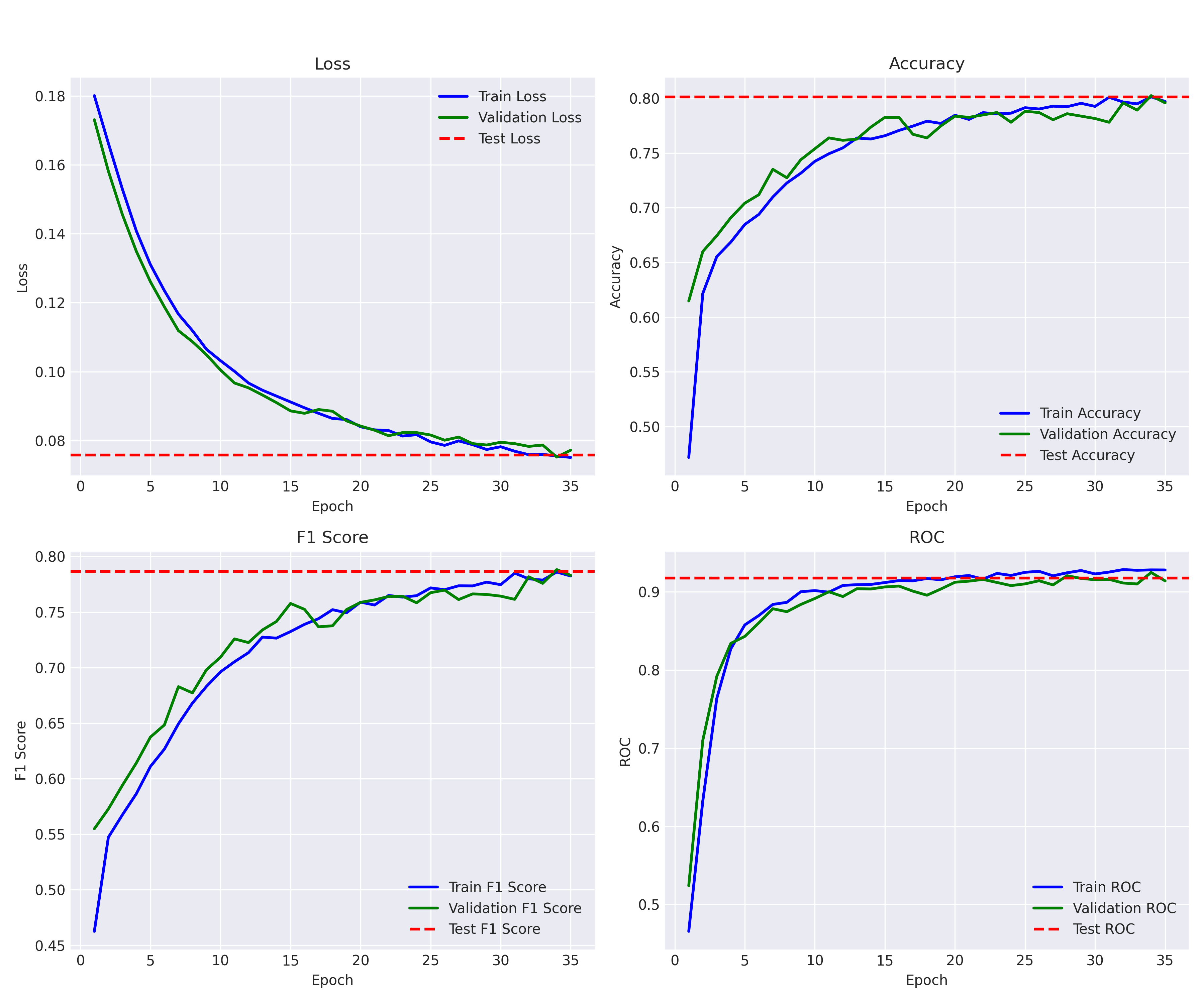}
    \caption{Model performance in noisy environment}\label{fig:noisy-simulation-plot}
\end{figure*}

\begin{table}[h!]
\caption{Hybrid Model Performance Results}\label{table:hybrid-model-results}
\begin{tabular}{@{}lllll@{}}
\toprule
\textbf{Environment} & MSE Loss & ROC & F-1 Score & Accuracy \\
\midrule
\textbf{Ideal} & 0.0488 & 0.9795 & 0.8786 & 0.8841 \\
\textbf{Noisy} & 0.0758 & 0.9177 & 0.7866 & 0.8013 \\
\botrule
\end{tabular}
\end{table}

The performance trends for the ideal and noisy simulations are visualized in Fig~\ref{fig:ideal-simulation-plot} and Fig~\ref{fig:noisy-simulation-plot} accordingly.

The evaluation metrics highlight the robustness of our hybrid recommendation system. These results demonstrate the effectiveness of combining classical and quantum-inspired methodologies for personalized recommendation tasks. The noisy environment test revealed some degradation in performance, but the hybrid model remained competitive and demonstrated resilience to noise.

\subsection{Quantitative Comparison}
In this section, we compare the performance of our model with several state-of-the-art approaches to collaborative filtering which are implemented using neural networks. The results presented below reflect key metrics including accuracy, F1-score, and loss values.

Recent advances in deep learning models, such as NeuMF \cite{19} and DeepCoNN \cite{20}, have shown great success in recommendation systems, achieving F1 scores typically ranging between 0.80 and 0.87.

Bobadilla et al.\cite{29} employed a classification-based neural network architecture to address collaborative filtering. Their model, trained over 80 epochs reported an accuracy of approximately 90\% on the MovieLens 1M dataset. In comparison, our model demonstrated the same accuracy trained over only 35 epochs, showcasing its ability to correctly classify user-item interactions with less computational power. Additionally, their reported F-1 Score of 0.55 is way below our models performance which is 0.87, explaining high performance in identifying user patterns with high precision. Even in the noisy configuration, our model could be considered superior in terms of F-1 Score.

Another study by Bobadilla et al.\cite{30} aimed to maximize the F1-score by leveraging deep learning for collaborative filtering. Their model achieved an F1-score of 0.45 which can be considered random, while our model produced a superior F1-score even in the noisy environment. This improvement in the F1-score indicates that our model strikes a better balance between precision and recall, which is crucial for producing high-quality recommendations in real-world applications.

Overall, our model consistently outperforms the models from these related works across various metrics, including accuracy, F1-score, and loss values, demonstrating its effectiveness in collaborative filtering tasks. The superior performance suggests that the architecture and training strategies we employed are well-suited for handling the complexities of recommendation systems and can be generalized to a variety of real-world applications.

Our hybrid quantum-classical model's performance is comparable to the state-of-the-art deep learning methods. However, our approach introduces quantum computation, which has the potential to accelerate both training and inference, particularly for high-dimensional datasets as was spotted. While deep learning models are adept at capturing complex user-item interactions, they often require significant computational resources. In contrast, the integration of quantum circuits in our model could offer computational efficiencies that reduce training time.

\subsection{Qualitative Comparison}
Herlocker et al. \cite{17} introduced a framework for evaluating collaborative filtering algorithms. They emphasized that metrics like Mean Squared Error (MSE) and accuracy alone do not adequately capture the overall effectiveness of recommendation systems. Instead, they proposed incorporating user-centric metrics such as serendipity, novelty, and user satisfaction, alongside ranking-based metrics like the Area Under the Receiver Operating Characteristic Curve (ROC AUC). In future evaluations of our model, we could focus on integrating these user-centric metrics, particularly novelty and serendipity, to provide a more comprehensive assessment of the system’s effectiveness in real-world applications.

Linden et al.'s work \cite{18} on item-to-item collaborative filtering, which powers Amazon's recommendation system, emphasizes scalability and real-time performance over traditional accuracy metrics. While Linden's approach is highly effective for large-scale, real-time recommendation tasks, their model does not report metrics such as Mean Squared Error (MSE) or F1 score, making direct comparison difficult. 

In contrast, our hybrid model, tested on the MovieLens dataset, has demonstrated high prediction accuracy but has yet to be evaluated in real-time, large-scale environments like those at Amazon. While Linden’s algorithm excels in rapid response times and scalability, our quantum-classical approach offers computational advantages, particularly in managing high-dimensional data through quantum circuits. 

Future work should focus on assessing our model's scalability in dynamic, large-scale applications to determine its practical utility in environments similar to Amazon's.

\section{Conclusion}\label{sec4}

We developed our quantum recommendation system with a hybrid architecture. This model utilizes an encoder to compress the dimensions of user vectors derived from the MovieLens dataset. Simultaneously, we used K-Means clustering to classify raw user vectors and extracted the cluster centers, which we then encoded and polarized to function as user archetypes. Next, we configured the quantum Hopfield associative memory to preserve the polarized patterns and retrieve them when new user vectors are received. Lastly, we added a Dense layer with a SoftMax activation function as the post-processing unit to determine the user category.

The model demonstrated satisfactory results compared to those achieved with traditional deep neural networks, positioning it as a viable option for future hybrid recommendation systems. 

\subsection{Future Works}
While our model shows strong quantitative performance, there is potential for future work beyond just accuracy-based evaluations. As Herlocker et al. suggest, incorporating metrics such as novelty and serendipity could provide deeper insights into the model’s ability to engage users over time. These metrics may highlight the system’s capability to generate surprising and engaging recommendations, potentially enhancing long-term user satisfaction and retention.

Although our quantum-enhanced model continues to utilize traditional classification methods, the results remain promising. Future research could focus on improving these techniques or integrating more advanced methods, such as Boltzmann Machines or Generative models, to further enhance the model’s flexibility and performance. Such advancements could enable the system to manage more complex datasets and offer even more personalized recommendations.

Testing the scalability of our model in large-scale, real-time applications remains a priority for future work. Additionally, broadening the evaluation criteria to include user engagement and long-term satisfaction would offer a more comprehensive understanding of the system's impact in practical settings.

Additionally, enhancing our model's fault tolerance can prepare it for implementation on real quantum hardware and evaluate its performance on a quantum device where there are errors present due to decoherence and environmental noises which are common in such hardware.

The application of quantum principles to cognitive models, as discussed in \cite{21}, suggests new possibilities for enhancing collaborative filtering systems. Quantum cognitive models could provide more nuanced representations of user preferences and behaviors, thereby improving the depth and accuracy of recommendations. Exploring these models may also yield new insights into group decision-making processes within social computing contexts. By leveraging quantum approaches in this manner, future recommendation systems could achieve a greater understanding of human cognition and behavior, leading to improved computational capabilities across various domains.
\section*{Acknowledgment} This work is supported by Iran Vice Presidency for Science, Technology and Knowledge-based Economy (ISTI), Grant no.140311240039

\bibliography{sn-bibliography}


\begin{thebibliography}{30}
\ifx \bisbn   \undefined \def \bisbn  #1{ISBN #1}\fi
\ifx \binits  \undefined \def \binits#1{#1}\fi
\ifx \bauthor  \undefined \def \bauthor#1{#1}\fi
\ifx \batitle  \undefined \def \batitle#1{#1}\fi
\ifx \bjtitle  \undefined \def \bjtitle#1{#1}\fi
\ifx \bvolume  \undefined \def \bvolume#1{\textbf{#1}}\fi
\ifx \byear  \undefined \def \byear#1{#1}\fi
\ifx \bissue  \undefined \def \bissue#1{#1}\fi
\ifx \bfpage  \undefined \def \bfpage#1{#1}\fi
\ifx \blpage  \undefined \def \blpage #1{#1}\fi
\ifx \burl  \undefined \def \burl#1{\textsf{#1}}\fi
\ifx \doiurl  \undefined \def \doiurl#1{\url{https://doi.org/#1}}\fi
\ifx \betal  \undefined \def \betal{\textit{et al.}}\fi
\ifx \binstitute  \undefined \def \binstitute#1{#1}\fi
\ifx \binstitutionaled  \undefined \def \binstitutionaled#1{#1}\fi
\ifx \bctitle  \undefined \def \bctitle#1{#1}\fi
\ifx \beditor  \undefined \def \beditor#1{#1}\fi
\ifx \bpublisher  \undefined \def \bpublisher#1{#1}\fi
\ifx \bbtitle  \undefined \def \bbtitle#1{#1}\fi
\ifx \bedition  \undefined \def \bedition#1{#1}\fi
\ifx \bseriesno  \undefined \def \bseriesno#1{#1}\fi
\ifx \blocation  \undefined \def \blocation#1{#1}\fi
\ifx \bsertitle  \undefined \def \bsertitle#1{#1}\fi
\ifx \bsnm \undefined \def \bsnm#1{#1}\fi
\ifx \bsuffix \undefined \def \bsuffix#1{#1}\fi
\ifx \bparticle \undefined \def \bparticle#1{#1}\fi
\ifx \barticle \undefined \def \barticle#1{#1}\fi
\bibcommenthead
\ifx \bconfdate \undefined \def \bconfdate #1{#1}\fi
\ifx \botherref \undefined \def \botherref #1{#1}\fi
\ifx \url \undefined \def \url#1{\textsf{#1}}\fi
\ifx \bchapter \undefined \def \bchapter#1{#1}\fi
\ifx \bbook \undefined \def \bbook#1{#1}\fi
\ifx \bcomment \undefined \def \bcomment#1{#1}\fi
\ifx \oauthor \undefined \def \oauthor#1{#1}\fi
\ifx \citeauthoryear \undefined \def \citeauthoryear#1{#1}\fi
\ifx \endbibitem  \undefined \def \endbibitem {}\fi
\ifx \bconflocation  \undefined \def \bconflocation#1{#1}\fi
\ifx \arxivurl  \undefined \def \arxivurl#1{\textsf{#1}}\fi
\csname PreBibitemsHook\endcsname

\bibitem[\protect\citeauthoryear{Biamonte et~al.}{2017}]{1}
\begin{barticle}
\bauthor{\bsnm{Biamonte}, \binits{J.}},
\bauthor{\bsnm{Wittek}, \binits{P.}},
\bauthor{\bsnm{Pancotti}, \binits{N.}},
\bauthor{\bsnm{Rebentrost}, \binits{P.}},
\bauthor{\bsnm{Wiebe}, \binits{N.}},
\bauthor{\bsnm{Lloyd}, \binits{S.}}:
\batitle{Quantum machine learning}.
\bjtitle{Nature}
\bvolume{549}(\bissue{7671}),
\bfpage{195}--\blpage{202}
(\byear{2017})
\doiurl{10.1038/nature23474}
\end{barticle}
\endbibitem

\bibitem[\protect\citeauthoryear{Peruzzo et~al.}{2014}]{2}
\begin{botherref}
\oauthor{\bsnm{Peruzzo}, \binits{A.}},
\oauthor{\bsnm{McClean}, \binits{J.}},
\oauthor{\bsnm{Shadbolt}, \binits{P.}},
\oauthor{\bsnm{Yung}, \binits{M.-H.}},
\oauthor{\bsnm{Zhou}, \binits{X.-Q.}},
\oauthor{\bsnm{Love}, \binits{P.J.}},
\oauthor{\bsnm{Aspuru-Guzik}, \binits{A.}},
\oauthor{\bsnm{O’Brien}, \binits{J.L.}}:
A variational eigenvalue solver on a photonic quantum processor.
Nature Communications
\textbf{5}(1)
(2014)
\doiurl{10.1038/ncomms5213}
\end{botherref}
\endbibitem

\bibitem[\protect\citeauthoryear{Farhi et~al.}{2014}]{3}
\begin{botherref}
\oauthor{\bsnm{Farhi}, \binits{E.}},
\oauthor{\bsnm{Goldstone}, \binits{J.}},
\oauthor{\bsnm{Gutmann}, \binits{S.}}:
A Quantum Approximate Optimization Algorithm
(2014).
\url{https://arxiv.org/abs/1411.4028}
\end{botherref}
\endbibitem

\bibitem[\protect\citeauthoryear{Duan et~al.}{2001}]{4}
\begin{barticle}
\bauthor{\bsnm{Duan}, \binits{L.-M.}},
\bauthor{\bsnm{Lukin}, \binits{M.D.}},
\bauthor{\bsnm{Cirac}, \binits{J.I.}},
\bauthor{\bsnm{Zoller}, \binits{P.}}:
\batitle{Long-distance quantum communication with atomic ensembles and linear optics}.
\bjtitle{Nature}
\bvolume{414}(\bissue{6862}),
\bfpage{413}--\blpage{418}
(\byear{2001})
\doiurl{10.1038/35106500}
\end{barticle}
\endbibitem

\bibitem[\protect\citeauthoryear{Schlosshauer}{2019}]{5}
\begin{barticle}
\bauthor{\bsnm{Schlosshauer}, \binits{M.}}:
\batitle{Quantum decoherence}.
\bjtitle{Physics Reports}
\bvolume{831},
\bfpage{1}--\blpage{57}
(\byear{2019})
\doiurl{10.1016/j.physrep.2019.10.001}
\end{barticle}
\endbibitem

\bibitem[\protect\citeauthoryear{Bussières et~al.}{2013}]{6}
\begin{barticle}
\bauthor{\bsnm{Bussières}, \binits{F.}},
\bauthor{\bsnm{Sangouard}, \binits{N.}},
\bauthor{\bsnm{Afzelius}, \binits{M.}},
\bauthor{\bsnm{Riedmatten}, \binits{H.}},
\bauthor{\bsnm{Simon}, \binits{C.}},
\bauthor{\bsnm{Tittel}, \binits{W.}}:
\batitle{Prospective applications of optical quantum memories}.
\bjtitle{Journal of Modern Optics}
\bvolume{60}(\bissue{18}),
\bfpage{1519}--\blpage{1537}
(\byear{2013})
\doiurl{10.1080/09500340.2013.856482}
\end{barticle}
\endbibitem

\bibitem[\protect\citeauthoryear{Hopfield}{1982}]{7}
\begin{barticle}
\bauthor{\bsnm{Hopfield}, \binits{J.}}:
\batitle{Neural networks and physical systems with emergent collective computational abilities}.
\bjtitle{Proceedings of the National Academy of Sciences of the United States of America}
\bvolume{79},
\bfpage{2554}--\blpage{8}
(\byear{1982})
\doiurl{10.1073/pnas.79.8.2554}
\end{barticle}
\endbibitem

\bibitem[\protect\citeauthoryear{Ventura and Martinez}{1998}]{8}
\begin{botherref}
\oauthor{\bsnm{Ventura}, \binits{D.}},
\oauthor{\bsnm{Martinez}, \binits{T.}}:
Quantum Associative Memory
(1998).
\url{https://arxiv.org/abs/quant-ph/9807053}
\end{botherref}
\endbibitem

\bibitem[\protect\citeauthoryear{Preskill}{2018}]{9}
\begin{barticle}
\bauthor{\bsnm{Preskill}, \binits{J.}}:
\batitle{Quantum computing in the nisq era and beyond}.
\bjtitle{Quantum}
\bvolume{2},
\bfpage{79}
(\byear{2018})
\doiurl{10.22331/q-2018-08-06-79}
\end{barticle}
\endbibitem

\bibitem[\protect\citeauthoryear{Deng et~al.}{2009}]{10}
\begin{bchapter}
\bauthor{\bsnm{Deng}, \binits{J.}},
\bauthor{\bsnm{Dong}, \binits{W.}},
\bauthor{\bsnm{Socher}, \binits{R.}},
\bauthor{\bsnm{Li}, \binits{L.-J.}},
\bauthor{\bsnm{Li}, \binits{K.}},
\bauthor{\bsnm{Li}, \binits{F.-F.}}:
\bctitle{Imagenet: a large-scale hierarchical image database},
pp. \bfpage{248}--\blpage{255}
(\byear{2009}).
\doiurl{10.1109/CVPR.2009.5206848}
\end{bchapter}
\endbibitem

\bibitem[\protect\citeauthoryear{Mari et~al.}{2020}]{11}
\begin{barticle}
\bauthor{\bsnm{Mari}, \binits{A.}},
\bauthor{\bsnm{Bromley}, \binits{T.R.}},
\bauthor{\bsnm{Izaac}, \binits{J.}},
\bauthor{\bsnm{Schuld}, \binits{M.}},
\bauthor{\bsnm{Killoran}, \binits{N.}}:
\batitle{Transfer learning in hybrid classical-quantum neural networks}.
\bjtitle{Quantum}
\bvolume{4},
\bfpage{340}
(\byear{2020})
\doiurl{10.22331/q-2020-10-09-340}
\end{barticle}
\endbibitem

\bibitem[\protect\citeauthoryear{Wang et~al.}{2019}]{12}
\begin{barticle}
\bauthor{\bsnm{Wang}, \binits{X.}},
\bauthor{\bsnm{Wang}, \binits{R.}},
\bauthor{\bsnm{Li}, \binits{D.}},
\bauthor{\bsnm{Adu-Gyamfi}, \binits{D.}},
\bauthor{\bsnm{Zhu}, \binits{Y.}}:
\batitle{Qcf: quantum collaborative filtering recommendation algorithm}.
\bjtitle{International Journal of Theoretical Physics}
\bvolume{58},
\bfpage{2235}--\blpage{2243}
(\byear{2019})
\end{barticle}
\endbibitem

\bibitem[\protect\citeauthoryear{Wang et~al.}{2021}]{13}
\begin{barticle}
\bauthor{\bsnm{Wang}, \binits{X.}},
\bauthor{\bsnm{Gu}, \binits{L.}},
\bauthor{\bsnm{Lee}, \binits{H.}},
\bauthor{\bsnm{Zhang}, \binits{G.}}:
\batitle{Quantum context-aware recommendation systems based on tensor singular value decomposition}.
\bjtitle{Quantum Information Processing}
\bvolume{20}(\bissue{5}),
\bfpage{1}--\blpage{32}
(\byear{2021})
\doiurl{10.1007/s11128-021-03131-y} .
\bcomment{Funding Information: This research is supported in part by Hong Kong Research Grant council (RGC) Grants (No. 15208418, No. 15203619, No. 15506619) and Shenzhen Fundamental Research Fund, China, under Grant No. JCYJ20190813165207290. Publisher Copyright: {\textcopyright} 2021, The Author(s), under exclusive licence to Springer Science+Business Media, LLC, part of Springer Nature.}
\end{barticle}
\endbibitem

\bibitem[\protect\citeauthoryear{Brahmachari et~al.}{2023}]{14}
\begin{botherref}
\oauthor{\bsnm{Brahmachari}, \binits{S.}},
\oauthor{\bsnm{Lumbreras}, \binits{J.}},
\oauthor{\bsnm{Tomamichel}, \binits{M.}}:
Quantum contextual bandits and recommender systems for quantum data
(2023).
\url{https://arxiv.org/abs/2301.13524}
\end{botherref}
\endbibitem

\bibitem[\protect\citeauthoryear{McClean et~al.}{2016}]{15}
\begin{barticle}
\bauthor{\bsnm{McClean}, \binits{J.R.}},
\bauthor{\bsnm{Romero}, \binits{J.}},
\bauthor{\bsnm{Babbush}, \binits{R.}},
\bauthor{\bsnm{Aspuru-Guzik}, \binits{A.}}:
\batitle{The theory of variational hybrid quantum-classical algorithms}.
\bjtitle{New Journal of Physics}
\bvolume{18}(\bissue{2}),
\bfpage{023023}
(\byear{2016})
\doiurl{10.1088/1367-2630/18/2/023023}
\end{barticle}
\endbibitem

\bibitem[\protect\citeauthoryear{Miller and Mukhopadhyay}{2021}]{16}
\begin{botherref}
\oauthor{\bsnm{Miller}, \binits{N.E.}},
\oauthor{\bsnm{Mukhopadhyay}, \binits{S.}}:
A quantum hopfield associative memory implemented on an actual quantum processor.
Scientific Reports
\textbf{11}(1)
(2021)
\doiurl{10.1038/s41598-021-02866-z}
\end{botherref}
\endbibitem

\bibitem[\protect\citeauthoryear{Mottonen et~al.}{2004}]{22}
\begin{botherref}
\oauthor{\bsnm{Mottonen}, \binits{M.}},
\oauthor{\bsnm{Vartiainen}, \binits{J.J.}},
\oauthor{\bsnm{Bergholm}, \binits{V.}},
\oauthor{\bsnm{Salomaa}, \binits{M.M.}}:
Transformation of quantum states using uniformly controlled rotations
(2004).
\url{https://arxiv.org/abs/quant-ph/0407010}
\end{botherref}
\endbibitem

\bibitem[\protect\citeauthoryear{Bergholm et~al.}{2022}]{24}
\begin{botherref}
\oauthor{\bsnm{Bergholm}, \binits{V.}},
\oauthor{\bsnm{Izaac}, \binits{J.}},
\oauthor{\bsnm{Schuld}, \binits{M.}},
\oauthor{\bsnm{Gogolin}, \binits{C.}},
\oauthor{\bsnm{Ahmed}, \binits{S.}},
\oauthor{\bsnm{Ajith}, \binits{V.}},
\oauthor{\bsnm{Alam}, \binits{M.S.}},
\oauthor{\bsnm{Alonso-Linaje}, \binits{G.}},
\oauthor{\bsnm{AkashNarayanan}, \binits{B.}},
\oauthor{\bsnm{Asadi}, \binits{A.}},
\oauthor{\bsnm{Arrazola}, \binits{J.M.}},
\oauthor{\bsnm{Azad}, \binits{U.}},
\oauthor{\bsnm{Banning}, \binits{S.}},
\oauthor{\bsnm{Blank}, \binits{C.}},
\oauthor{\bsnm{Bromley}, \binits{T.R.}},
\oauthor{\bsnm{Cordier}, \binits{B.A.}},
\oauthor{\bsnm{Ceroni}, \binits{J.}},
\oauthor{\bsnm{Delgado}, \binits{A.}},
\oauthor{\bsnm{Matteo}, \binits{O.D.}},
\oauthor{\bsnm{Dusko}, \binits{A.}},
\oauthor{\bsnm{Garg}, \binits{T.}},
\oauthor{\bsnm{Guala}, \binits{D.}},
\oauthor{\bsnm{Hayes}, \binits{A.}},
\oauthor{\bsnm{Hill}, \binits{R.}},
\oauthor{\bsnm{Ijaz}, \binits{A.}},
\oauthor{\bsnm{Isacsson}, \binits{T.}},
\oauthor{\bsnm{Ittah}, \binits{D.}},
\oauthor{\bsnm{Jahangiri}, \binits{S.}},
\oauthor{\bsnm{Jain}, \binits{P.}},
\oauthor{\bsnm{Jiang}, \binits{E.}},
\oauthor{\bsnm{Khandelwal}, \binits{A.}},
\oauthor{\bsnm{Kottmann}, \binits{K.}},
\oauthor{\bsnm{Lang}, \binits{R.A.}},
\oauthor{\bsnm{Lee}, \binits{C.}},
\oauthor{\bsnm{Loke}, \binits{T.}},
\oauthor{\bsnm{Lowe}, \binits{A.}},
\oauthor{\bsnm{McKiernan}, \binits{K.}},
\oauthor{\bsnm{Meyer}, \binits{J.J.}},
\oauthor{\bsnm{Montañez-Barrera}, \binits{J.A.}},
\oauthor{\bsnm{Moyard}, \binits{R.}},
\oauthor{\bsnm{Niu}, \binits{Z.}},
\oauthor{\bsnm{O'Riordan}, \binits{L.J.}},
\oauthor{\bsnm{Oud}, \binits{S.}},
\oauthor{\bsnm{Panigrahi}, \binits{A.}},
\oauthor{\bsnm{Park}, \binits{C.-Y.}},
\oauthor{\bsnm{Polatajko}, \binits{D.}},
\oauthor{\bsnm{Quesada}, \binits{N.}},
\oauthor{\bsnm{Roberts}, \binits{C.}},
\oauthor{\bsnm{Sá}, \binits{N.}},
\oauthor{\bsnm{Schoch}, \binits{I.}},
\oauthor{\bsnm{Shi}, \binits{B.}},
\oauthor{\bsnm{Shu}, \binits{S.}},
\oauthor{\bsnm{Sim}, \binits{S.}},
\oauthor{\bsnm{Singh}, \binits{A.}},
\oauthor{\bsnm{Strandberg}, \binits{I.}},
\oauthor{\bsnm{Soni}, \binits{J.}},
\oauthor{\bsnm{Száva}, \binits{A.}},
\oauthor{\bsnm{Thabet}, \binits{S.}},
\oauthor{\bsnm{Vargas-Hernández}, \binits{R.A.}},
\oauthor{\bsnm{Vincent}, \binits{T.}},
\oauthor{\bsnm{Vitucci}, \binits{N.}},
\oauthor{\bsnm{Weber}, \binits{M.}},
\oauthor{\bsnm{Wierichs}, \binits{D.}},
\oauthor{\bsnm{Wiersema}, \binits{R.}},
\oauthor{\bsnm{Willmann}, \binits{M.}},
\oauthor{\bsnm{Wong}, \binits{V.}},
\oauthor{\bsnm{Zhang}, \binits{S.}},
\oauthor{\bsnm{Killoran}, \binits{N.}}:
PennyLane: Automatic differentiation of hybrid quantum-classical computations
(2022).
\url{https://arxiv.org/abs/1811.04968}
\end{botherref}
\endbibitem

\bibitem[\protect\citeauthoryear{Cao et~al.}{2017}]{23}
\begin{botherref}
\oauthor{\bsnm{Cao}, \binits{Y.}},
\oauthor{\bsnm{Guerreschi}, \binits{G.G.}},
\oauthor{\bsnm{Aspuru-Guzik}, \binits{A.}}:
Quantum Neuron: an elementary building block for machine learning on quantum computers
(2017).
\url{https://arxiv.org/abs/1711.11240}
\end{botherref}
\endbibitem

\bibitem[\protect\citeauthoryear{Aseguinolaza et~al.}{2024}]{28}
\begin{botherref}
\oauthor{\bsnm{Aseguinolaza}, \binits{U.}},
\oauthor{\bsnm{Sobrino}, \binits{N.}},
\oauthor{\bsnm{Sobrino}, \binits{G.}},
\oauthor{\bsnm{Jornet-Somoza}, \binits{J.}},
\oauthor{\bsnm{Borge}, \binits{J.}}:
Error estimation in current noisy quantum computers.
Quantum Information Processing
\textbf{23}(5)
(2024)
\doiurl{10.1007/s11128-024-04384-z}
\end{botherref}
\endbibitem

\bibitem[\protect\citeauthoryear{IBM}{2024}]{27}
\begin{botherref}
\oauthor{\bsnm{IBM}}:
Qiskit AER Noise Models.
\url{https://qiskit.github.io/qiskit-aer/apidocs/aer_noise.html}
Accessed 2024/12/10
\end{botherref}
\endbibitem

\bibitem[\protect\citeauthoryear{Javadi-Abhari et~al.}{2024}]{25}
\begin{botherref}
\oauthor{\bsnm{Javadi-Abhari}, \binits{A.}},
\oauthor{\bsnm{Treinish}, \binits{M.}},
\oauthor{\bsnm{Krsulich}, \binits{K.}},
\oauthor{\bsnm{Wood}, \binits{C.J.}},
\oauthor{\bsnm{Lishman}, \binits{J.}},
\oauthor{\bsnm{Gacon}, \binits{J.}},
\oauthor{\bsnm{Martiel}, \binits{S.}},
\oauthor{\bsnm{Nation}, \binits{P.D.}},
\oauthor{\bsnm{Bishop}, \binits{L.S.}},
\oauthor{\bsnm{Cross}, \binits{A.W.}},
\oauthor{\bsnm{Johnson}, \binits{B.R.}},
\oauthor{\bsnm{Gambetta}, \binits{J.M.}}:
Quantum computing with {Q}iskit
(2024).
\doiurl{10.48550/arXiv.2405.08810}
\end{botherref}
\endbibitem

\bibitem[\protect\citeauthoryear{IBM}{2024}]{26}
\begin{botherref}
\oauthor{\bsnm{IBM}}:
Qiskit AER Simulator Provider.
\url{https://qiskit.github.io/qiskit-aer/apidocs/aer_provider.html}
Accessed 2024/12/10
\end{botherref}
\endbibitem

\bibitem[\protect\citeauthoryear{He et~al.}{2017}]{19}
\begin{botherref}
\oauthor{\bsnm{He}, \binits{X.}},
\oauthor{\bsnm{Liao}, \binits{L.}},
\oauthor{\bsnm{Zhang}, \binits{H.}},
\oauthor{\bsnm{Nie}, \binits{L.}},
\oauthor{\bsnm{Hu}, \binits{X.}},
\oauthor{\bsnm{Chua}, \binits{T.-S.}}:
Neural Collaborative Filtering
(2017).
\url{https://arxiv.org/abs/1708.05031}
\end{botherref}
\endbibitem

\bibitem[\protect\citeauthoryear{Zheng et~al.}{2017}]{20}
\begin{botherref}
\oauthor{\bsnm{Zheng}, \binits{L.}},
\oauthor{\bsnm{Noroozi}, \binits{V.}},
\oauthor{\bsnm{Yu}, \binits{P.S.}}:
Joint Deep Modeling of Users and Items Using Reviews for Recommendation
(2017).
\url{https://arxiv.org/abs/1701.04783}
\end{botherref}
\endbibitem

\bibitem[\protect\citeauthoryear{Bobadilla et~al.}{2020a}]{29}
\begin{barticle}
\bauthor{\bsnm{Bobadilla}, \binits{J.}},
\bauthor{\bsnm{Ortega}, \binits{F.}},
\bauthor{\bsnm{Gutiérrez}, \binits{A.}},
\bauthor{\bsnm{Alonso}, \binits{S.}}:
\batitle{Classification-based deep neural network architecture for collaborative filtering recommender systems}.
\bjtitle{International Journal of Interactive Multimedia and Artificial Intelligence}
\bvolume{6}(\bissue{1}),
\bfpage{68}--\blpage{77}
(\byear{2020})
\doiurl{10.9781/ijimai.2020.02.006}
\end{barticle}
\endbibitem

\bibitem[\protect\citeauthoryear{Bobadilla et~al.}{2020b}]{30}
\begin{botherref}
\oauthor{\bsnm{Bobadilla}, \binits{J.}},
\oauthor{\bsnm{Alonso}, \binits{S.}},
\oauthor{\bsnm{Hernando}, \binits{A.}}:
Deep learning architecture for collaborative filtering recommender systems.
Applied Sciences
\textbf{10}(7)
(2020)
\doiurl{10.3390/app10072441}
\end{botherref}
\endbibitem

\bibitem[\protect\citeauthoryear{Herlocker et~al.}{2004}]{17}
\begin{barticle}
\bauthor{\bsnm{Herlocker}, \binits{J.L.}},
\bauthor{\bsnm{Konstan}, \binits{J.A.}},
\bauthor{\bsnm{Terveen}, \binits{L.G.}},
\bauthor{\bsnm{Riedl}, \binits{J.T.}}:
\batitle{Evaluating collaborative filtering recommender systems}.
\bjtitle{ACM Trans. Inf. Syst.}
\bvolume{22}(\bissue{1}),
\bfpage{5}--\blpage{53}
(\byear{2004})
\doiurl{10.1145/963770.963772}
\end{barticle}
\endbibitem

\bibitem[\protect\citeauthoryear{Linden et~al.}{2003}]{18}
\begin{barticle}
\bauthor{\bsnm{Linden}, \binits{G.}},
\bauthor{\bsnm{Smith}, \binits{B.}},
\bauthor{\bsnm{York}, \binits{J.}}:
\batitle{Amazon.com recommendations: item-to-item collaborative filtering}.
\bjtitle{IEEE Internet Computing}
\bvolume{7}(\bissue{1}),
\bfpage{76}--\blpage{80}
(\byear{2003})
\doiurl{10.1109/MIC.2003.1167344}
\end{barticle}
\endbibitem

\bibitem[\protect\citeauthoryear{Busemeyer and Bruza}{2012}]{21}
\begin{bbook}
\bauthor{\bsnm{Busemeyer}, \binits{J.R.}},
\bauthor{\bsnm{Bruza}, \binits{P.D.}}:
\bbtitle{Quantum Models of Cognition and Decision}.
\bpublisher{Cambridge University Press}, \blocation{???}
(\byear{2012})
\end{bbook}
\endbibitem

\end{thebibliography}

\end{document}